# Quantitative identification of technological discontinuities using simulation modeling

Hyunseok Park[1,3]* and Christopher L. Magee[2,3]

[1] Department of Information System, Hanyang University, Seoul, Republic of Korea

[2] SUTD-MIT International Design Center, Massachusetts Institute of Technology (MIT), Cambridge, Massachusetts, United States

[3] Institute of Data, Systems, and Society, Massachusetts Institute of Technology (MIT), Cambridge, Massachusetts, United States

Email: Hyunseok Park (hp@hanyang.ac.kr) and Christopher L. Magee (cmagee@mit.edu)

*Corresponding author: Hyunseok Park (hp@hanyang.ac.kr)

## Abstract

The aim of this paper is to develop and test metrics to quantitatively identify technological discontinuities in a knowledge network. We developed five metrics based on innovation theories and tested the metrics by a simulation model-based knowledge network and hypothetically designed discontinuity. The designed discontinuity is modeled as a node which combines two different knowledge streams and whose knowledge is dominantly persistent in the knowledge network. The performances of the proposed metrics were evaluated by how well the metrics can distinguish the designed discontinuity from other nodes on the knowledge network. The simulation results show that the *persistence times # of converging main paths* provides the best performance in identifying the designed discontinuity: the designed discontinuity was identified as one of the top 3 patents with 96~99% probability by Metric 5 and it is, according to the size of a domain, 12~34% better than the performance of the second best metric. Beyond the simulation analysis, we tested the metrics using a patent set representative of the Magnetic information storage domain. The three representative patents associated with a well-known breakthrough technology in the domain, the giant magneto-resistance (GMR) spin valve sensor, were selected based on the qualitative studies, and the metrics were tested by how well the metrics identify the selected patents as top-ranked patents. The empirical results fully support the simulation results and therefore the *persistence times # of converging main paths* is recommended for identifying technological discontinuities for any technology.

## 1. Introduction

Technological discontinuities[1] have been hypothesized to be a major factor in continual technological progress (Tushman and Anderson, 1986) and in market competitions when they enable surpassing incumbent approaches, obsoleting prevailing dominant designs and upsetting market structure (Ehrnberg, 1995; Hill and Utterback, 1980; Hill and Rothaermel, 2003; Utterback and Abernathy, 1975; Utterback and Suarez, 1993). Reproducible identification of technological discontinuities provides an important objective for both academic research and managerial practice. In innovation studies, objective identification of technological discontinuities could enable an increase in empirical understanding of technological progress and performance dynamics. For firms, the ability to anticipate or at least recognize technological discontinuities is not only a critical tool for maintaining or strengthening a firm's competitive advantage, but also potentially a means for overcoming competitive disadvantages (Danneels, 2004; Hill and Rothaermel, 2003; Rothaermel, 2000).

[1] The technological discontinuities arise in knowledge networks and thus are also referred to as knowledge discontinuities in this paper.

Although prior research has suggested approaches to identify technological discontinuities (Anderson and Tushman, 1990; Hoisl et al., 2015; Martinelli and Nomaler, 2014; Tushman and Anderson, 1986), they have not been made operational and objective. In this paper, we propose a method for quantitative identification of technological discontinuities in a technological domain (TD)[2]. Specifically, we develop metrics to quantitatively identify technological discontinuities in a knowledge network, represented here by a patent citation network.

Patents are suitable data for this work as they contain most of the technological knowledge in a specific TD (Abraham and Moitra, 2001; Griliches, 1990; Hall et al., 2001; Von Wartburg et al., 2005) and each patent differs in some way from prior art and so has a degree of novelty, i.e. every patent introduces some discontinuity in knowledge regardless of its strength so we are looking for a method to identify the strongest discontinuities. Patent citations are evidence that the knowledge disclosed in the cited patents are relevant to the citing patent and simplistically the citing patent can be considered as the novel inventive knowledge created by the combination of knowledge in the cited patents. In a patent citation network, nodes are patents and links are citations and paths in such networks can represent the accumulation of knowledge in technological trajectories (Fontana et al., 2009; Martinelli, 2012; Mina et al., 2007; Verspagen, 2007). Therefore, the technological discontinuities in a patent citation network are the patents that have strongly different underlying knowledge, in other words, a different technological paradigm (Dosi, 1982; Martinelli, 2012; Nelson and Winter, 1977; Verspagen, 2007), from previous patents in their technological lineage.

The metrics were deduced from existing innovation theory as covered in Section 2. To test the metrics, we developed a simulation model-based patent citation network containing a hypothetically designed discontinuity (see Section 3). The performance of the metrics is evaluated by how well the metrics distinguish the designed discontinuity from other nodes in a patent citation network simulation. The simulation results show that Metric 5 (*persistence times the number of converging main paths*) provides the best performance in identifying the designed discontinuity; the designed discontinuity was identified as one of the top 3 patents with 96~99% probability by Metric 5 and it is, according to the size of a domain, 12~34% better than the performance of the second best metric.

Beyond the simulation analysis, we also conducted an empirical analysis using a patent set representative of Magnetic information storage. Based upon prior deep qualitative studies for this domain, we first identified the three patents that embodied an important technological discontinuity in the Magnetic information storage domain, a giant magneto-resistance (GMR) spin-valve sensor. The empirical tests show that the best metric (Metric 5) identified in the simulation also has good performance in the case study. Therefore, based on both simulation and empirical tests, we concluded that *persistence times the number of converging main paths* is the most useful metric for identifying technological discontinuities among our five metrics.

[2] We adopted Magee, C. L., et al., (2016)'s definition for a technological domain: The set of artifacts that fulfill a specific generic function utilizing a particular, recognizable body of knowledge.

The rest of this paper is structured as follows: Section 2 reviews the literature on technological discontinuities and develops the identification metrics based on theoretical concepts, Section 3 describes the details of the simulation modeling, Section 4 provides and discusses the simulation results, Section 5 discusses the results of the magnetic information storage empirical analysis and Section 6 presents the conclusions of this research.

## 2. Technological discontinuities

This section reviews the previous literature on technological discontinuities first noting that the literature contains different perspectives to explain technological discontinuities. Therefore, we first derive main concepts from the theories, then formulate specific metrics to identify technological discontinuities based on the concepts.

First, there is a widely accepted agreement that rapid improvement of a TD is dominantly affected by very important inventions (Tushman and Anderson, 1986). These inventions are discussed theoretically using terms such as breakthrough (Sharpe et al., 2013), discontinuous (Tushman and Anderson, 1986), non-incremental (Nemet, 2009), radical (Ettlie et al., 1984), or disruptive innovation (Christensen, 1997; Kassicieh et al., 2002). Although different innovation theories have different criteria for determining the importance of inventions and/or related sets of inventions, many technologically very important inventions are agreed upon regardless of the different theoretical perspectives. For instance, Blue Light Emission Diode (LED) is a well known very important invention in lighting technology and is categorized as a very important invention by each innovation theory. Therefore, our first concept for technological discontinuities is as follows:

***Concept A****: technological discontinuities are the inventions that are technologically important in a technological domain.*

Second, one theoretical approach to describe technological evolution is a cyclical model: the emergence of technological discontinuities breaks stable periods initiated by previous discontinuities (a so-called dominant design), and then a new dominant design incorporating the latest discontinuity is established and leads again to a new stable period of incremental changes (Abernathy and Utterback, 1978; Anderson and Tushman, 1990; Kaplan, 1999; Munir, 2003; Sahal, 1981; Tushman, 1997; Tushman and Nelson, 1990; Utterback, 1994). A related but somewhat different framework – technological paradigms and trajectories (Dosi, 1982) – focuses on knowledge rather than artifacts but describes a very similar cyclical pattern. The emergence of a new technological paradigm, which is a model and pattern of solution of selected technological problems, makes the existing paradigm obsolete, less useful or even useless knowledge and provides the underlying framework for the incremental problem solving in technological trajectories (Dosi, 1982). Tushman et al. (1997) suggest that discontinuous innovation breaks with the past stream to create new technologies, processes, and organizational architecture. Utterback (1994) described the discontinuous innovation as involving discontinuities or radical innovation that allows entire industries to emerge or disappear. Ahuja and Lampert (2001) noted that breakthrough inventions 'serve as the basis of new technological trajectories and paradigms and are an important part of the process of creative destruction in which extant techniques and approaches are replaced by new technologies and products'. In summary, the specific concept is:

***Concept B**: technological discontinuities arise from the inventions related to a new technological paradigm or knowledge not used in the previous paradigm.*

Third, most innovation theories agree that there is no entirely new technological knowledge: the recombination and reconfiguration of existing knowledge is the principal mechanism of invention or generation of new technological knowledge (Basnet and Magee, 2016; Dahlin and Behrens, 2005; Della Malva and Riccaboni, 2014; Fleming, 2001; Fleming and Sorenson, 2001; Gilfillan, 1935; Henderson and Clark, 1990; Nelson and Winter, 1982; Penrose, 1959; Schilling and Green, 2011; Schumpeter, 1934; Usher, 1954; Uzzi et al., 2013; Weitzman, 1998; Youn et al., 2015). Most recombination is based on combining local or familiar knowledge (Cyert and March, 1963; Dosi, 1988; Stuart and Podolny, 1996) and this type of combination is likely to deliver incremental improvements (Fleming, 2001). Whereas, the combinations of unconventional or unfamiliar knowledge are more likely to generate high novelty and are regarded as the foundation of breakthrough or radical inventions (Fleming, 2001; Simonton, 1999). Recent empirical studies indicate that the unconventional combinations are the sources of high impact knowledge on radical innovation in technologies (Ahuja and Lampert, 2001; Della Malva and Riccaboni, 2014; Kelley et al., 2013; Schoenmakers and Duysters, 2010; Singh and Fleming, 2010; Strumsky and Lobo, 2015) and such combinations are also important in science (Schilling and Green, 2011; Uzzi et al., 2013). Based on this stream of research, the third concept is:

***Concept C**: technological discontinuities are the inventions generated by combination of unconventional knowledge.*

Fourth, some papers define technological discontinuities as innovations that provide dramatically high advantages in price or performance, so any efforts to increase in scale, efficiency, or design cannot make the older technologies be competitive (Anderson and Tushman, 1990; Mensch, 1979; Rice et al., 1998; Sahal, 1981; Schumpeter, 1942; Tushman and Anderson, 1986). Schumpeter mentioned "command a decisive cost or quality advantage and that strike not at the margins of the profits and the outputs of the existing firms, but at their foundations and their very lives". Tushman and Anderson (1986) and Anderson and Tushman (1990) defined technological discontinuities as innovations that depart significantly from the general underlying knowledge of continuous incremental innovation, and provided examples of discontinuities in the Cement, Glass, Airline, and Microcomputer industries. Rice et al. (1998) defined discontinuous innovations as 'game changers', which have the potential for a 5-10 times improvement in performance and for a 30-50% percent reduction in cost compared to existing products, such as GE's digital X-ray, GM's hybrid vehicle, IBM's silicon-germanium devices, and so forth. However, recent empirical quantitative technological change studies show empirical results that performance of every TD continually improves exponentially with time from the long term perspective (Benson and Magee, 2015a; Koh and Magee, 2006, 2008; Magee et al., 2016). In this last paper, it is concluded that previous research does not provide clear empirical and quantitative evidences on dramatic discontinuous improvements in price or performance. For example, although above mentioned literatures showed empirical examples on dramatic performance improvement in several TDs, this is not unexpected in rapidly improving domains where the possibility of many missing data between data points exist. Thus, no known results demonstrate statistically reliable discontinuities in performance. Therefore, it seems inappropriate to link price or performance dynamics of a TD to specific technological discontinuities.

## 3. Method

We developed five metrics and a simulation model to test the metrics. In this section, we provide a detailed description of the metrics and the simulation model.

### 3.1. Metrics

The metrics for quantitative identification and evaluation of knowledge discontinuities in a patent citation network are developed based on the concepts in Section 2. Each of these metrics is proposed as a measure of the strength of the discontinuity associated with a given patent. The specifics for each metric are described below.

***Concept A***

We developed three metrics from *Concept A* concerned with the importance of patents:

There has been significant effort to evaluate the economical or technological value of a single patent (Carpenter et al., 1981; Hall et al., 2005; Harhoff et al., 1999; Harhoff et al., 2003; Jaffe and Trajtenberg, 2002; Trajtenberg, 1990) using forward citation information. Since the distribution of patent values is highly skewed, most patents have a low value of forward citations but a very few patents are highly cited. There has been extensive work showing that these highly cited patents are in fact valuable (Fischer and Leidinger, 2014; Harhoff et al., 1999; Jaffe et al., 2000) and important in technological progress (Benson and Magee, 2015a; Girifalco, 1991; Sahal, 1981; Trajtenberg, 1990; Tushman and Anderson, 1986). Many indicators have been suggested and most of them utilize patent citations as signaling importance. Patent citation-based indicators can be broadly divided into two types: local citation count-based and global citation structure-based approaches. We use the local citation count-based approach partly because we want to test our simulation models and these only describe citations within a domain. We measure the technological importance of the patent *i* ($Pat_i$) in the TD by the number of forward citations for $Pat_i$ within the TD.

$$Metric\ 1 = FWDCIT,$$

where $FWDCIT$ is the number of forward citations to the patent of interest by other patents in the TD.

Global citation structure-based approaches measure both direct and indirect citation relationships between patents, and we adopted the genetic knowledge inheritance algorithm, suggested by Martinelli and Nomaler (2014), to measure how much knowledge of a patent remains in or contributes to later patents, called knowledge persistence of a patent. If a patent has a high persistence value, i.e. is a high persistence patent (HPP), its inventive knowledge dominantly contributes as an essential ingredient to descendent patents in the TD. The knowledge persistence of patents is taken as a proxy for knowledge dominance or the persistence of their technological contribution in a TD. Since the actual persistence value can be affected by the patent application or grant year, i.e. older patents have a higher chance to get high persistence than recent patents, the effect of time must be considered. Metric 2 associates the strength of the knowledge discontinuity for a patent as its persistence value (Metric 2).

$$Metric\ 2 = P,$$

where $P$ is persistence value of a patent.

It is plausible that knowledge discontinuities have not only dominantly important technological knowledge in TD, but also direct impacts on other patents in the TD. However, all highly cited patents do not have high persistence, and vice versa, because knowledge persistence and forward citations have only modest correlation[3]. Therefore, we developed one more metric (Metric 3) that can identify patents having *both* high persistence and forward citation frequency, and Metric 3 is defined as follows:

$$Metric\ 3 = FWDCIT \cdot P.$$

***Concept B***

The metric we derive from *Concept B* – technological discontinuities as new technological paradigms or dominant designs have different underlying knowledge from the past knowledge stream – characterizes discontinuous points on technological trajectories. Main path analysis of patent citation networks has been relatively widely used for empirically identifying and visualizing technological trajectories (Park and Magee, 2016)[4]. Therefore, a discontinuous, or weak, link between patents on the main paths can be a signal for a knowledge discontinuity. On the main paths, if a HPP is not directly connected to any prior HPPs, it is interpreted that the previous knowledge stream is disconnected and replaced by a new knowledge stream (Fig 1). The disconnected HPP can be identified as a knowledge discontinuity and a metric which infers the degree of discontinuity based on the minimum persistence value between HPPs is metric 4 calculated as follows:

$$Metric\ 4 = \frac{1}{MP} - 1,$$

where $MP$ is the minimum persistence value between HPPs.

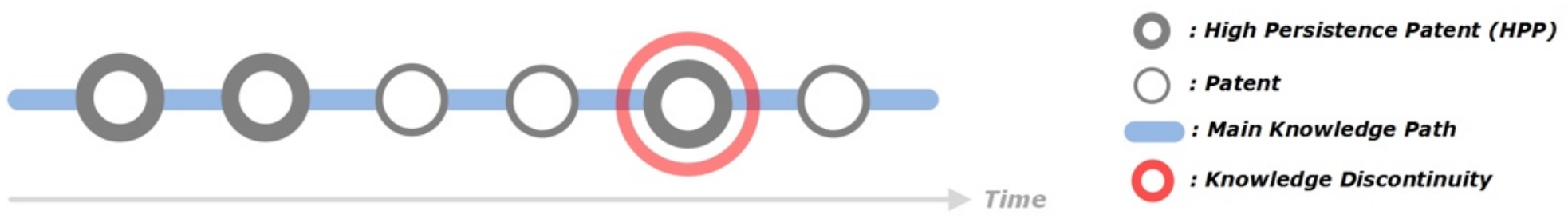

Fig 1. Discontinuity on main path.

***Concept C*** – combinations of unconventional knowledge

The network of main paths of a TD generally consists of a multiple number of main paths, because there exists different approaches and components to fulfill a specific function of the TD and they each have their own somewhat independent developmental trajectories. In the technological evolutionary process, paths, i.e. knowledge streams, are frequently converged or diverged to create better engineering solutions. Therefore, path convergence is a combination of two previously un-related knowledge streams to form a new stream and can be interpreted as combining unconventional knowledge. A patent which combines unimportant incoming different knowledge streams is not likely to be important but a patent is more likely to be a technological discontinuity if its new inventive knowledge is technologically or dominantly significant in the TD (Fig 2). We defined the technological discontinuities as patents that converge different main paths and have high persistence value in the TD. However, a high frequency of backward citation would tend to accidentally touch on many converging main paths. In order to avoid the high backward citation effect, we normalized the size of the converging main paths by dividing by (*1+the number of in-domain backward citations from the patent of interest + PATH*) and therefore the strength of the discontinuity is calculated as follows (Metric 5):

$$Metric\ 5 = Normalized\ size\ of\ convering\ main\ paths \times P = \frac{PATH}{1+BWDCIT+PATH} \cdot P,$$

where PATH is the number of converging main paths on a specific node, if a patent is the first node of a main path, we set the PATH of the patent as one; BWDCIT is the number of backward citations to prior patents in TD by the patent of interest.

---

[3] We tested correlations between persistence value and forward citations using top 50 persistence and forward citation patents and r=0.221 (p<0.0001) from 100 simulations of a patent citation model composed of 1000 patents and r=-0.215 (p<0.0001) from a patent set for the magnetic information storage.

[4] We adopted the knowledge persistence-based main path approach, developed by Park and Magee (2016), which can include all HPPs in a citation network.

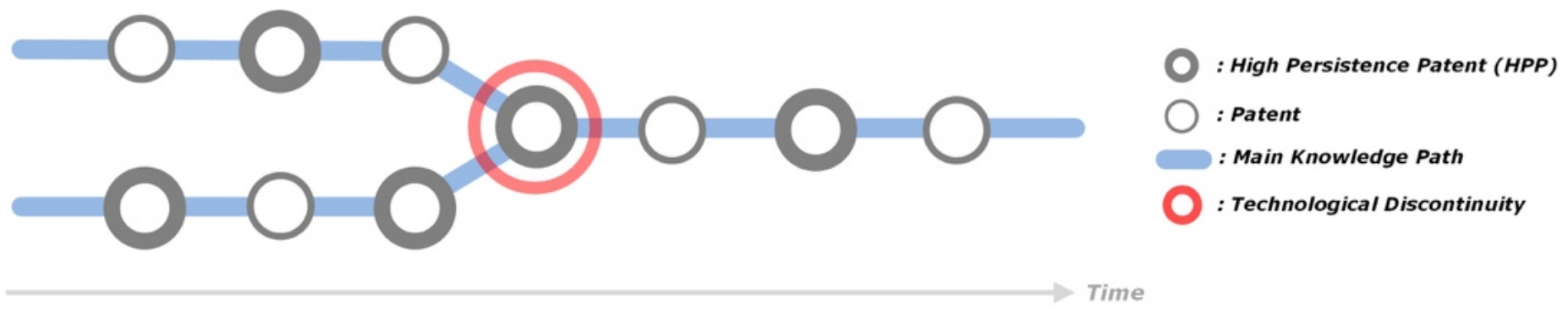

Fig 2. Convergence of main paths.

### 3.2. Simulation model

#### 3.2.1. Modeling of patent citation networks

The patent system has generic characteristics that lead to similarities in all citation networks: we use these characteristics to develop our patent citation network model. The generic characteristics involve: scale growth, average citations, citation frequency, and citation lag.

- *Scale growth*: the number of patents tends to grow by a constant 5% (actually 4.55%) per year, i.e. exponential growth, and this 5% was calculated using all granted patents in the United States Patent and Trademark Office (USPTO) database from 1981 to 2014[5].
- *Average citation:* on average, patents have approximately three forward citations, backward citations also, to other patents in the same TD[6].
- *Citation frequency:* patent forward citations follow a power law and thus some patents receive many more forward citations than most (Silverberg and Verspagen, 2007).

5 http://www.uspto.gov/web/offices/ac/ido/oeip/taf/us_stat.htm
6 We adopted the patent search method, suggested by Benson and Magee (2013, 2015b) to identify a set of patents for a specific TD, and the average number of citations for six TDs (MRI, Fuel cell, LED, Magnetic information storage, Solar PV, and 3D printing) is 3.08.

- *Citation lag:* the forward citations a patent receives tend to peak after 3-5 years of publication (Hall et al., 2001).

We also select the number of patents we want to have in the model (the number of patents, *n,* varied from 600 to 30,000 reflecting possible real world domains) and the number of years we want to cover (30 was selected for all cases). The procedure for modeling a patent citation network, which satisfies the aforementioned properties, is as follows:

1) *Establishing the number of patents for each year,*
2) *Establishing the number of forward citations for each patent,*
3) *Assigning a specific year for each patent,*
4) *Selecting specific forward citations for each patent.*

First, the number of patents for each year is determined (Fig 3). Since the number of patents increases at an average rate of 5% per year, we assume that the yearly patent count, $y$, follows an exponential trend with time, $t$, given by $y = a \cdot e^{0.05t} + b$, where $a$ and $b$ are empirical fitting parameters. In order to set an initial time with zero patents, the values of $a$ and $b$ are set to $a$=1 and $b$=-1, so that $y$=0 at $x$=0. The values of $y$ for each year are normalized by the total value of $y$ to obtain a fractional value ranging between 0 and 1. Then, the patent generation process is executed by a simple Monte Carlo routine where a random number generator is used to generate $n$ values following a uniform distribution between 0 and 1. Each of $n$ random values is compared to the relative proportions of $y$ between 0 and 1 and assigned to one of the year-intervals.

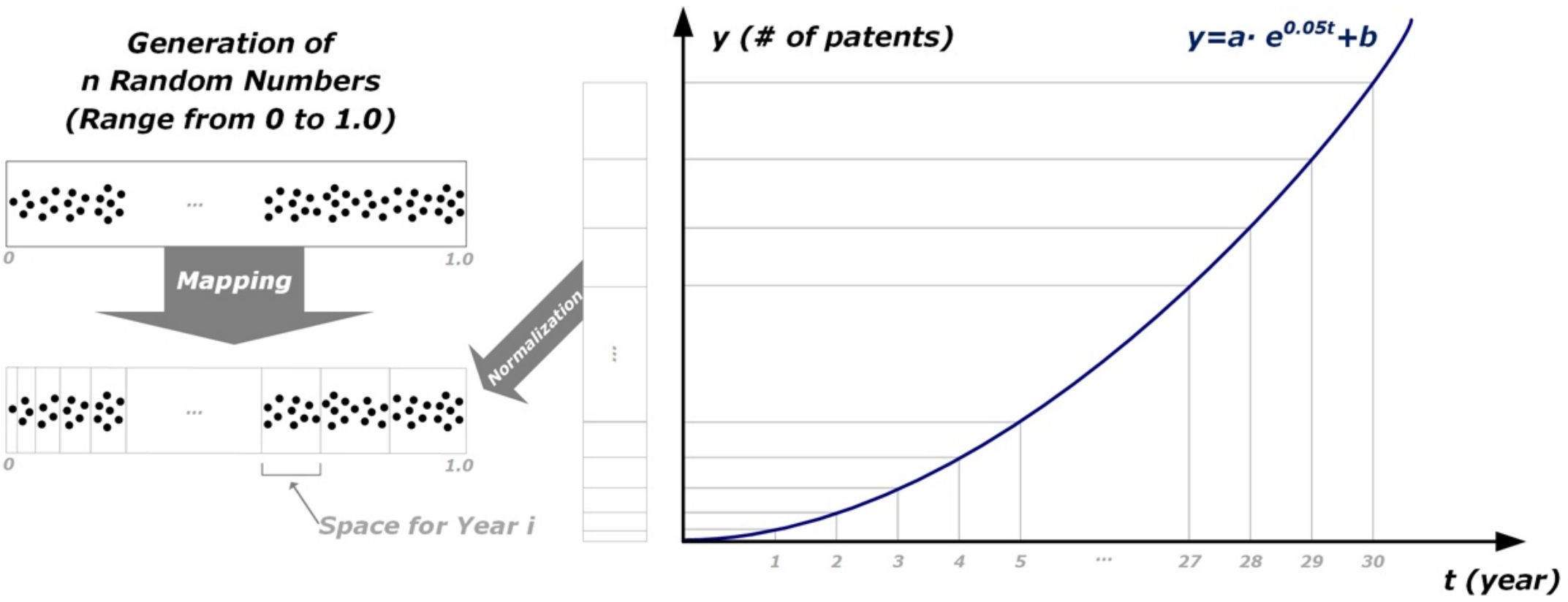

Fig 3. Number of assigned patents to each year

Second, the number of forward citations for each patent is determined (Fig 4). The distribution of the number of forward citations, $F$, follows a power law as a function of the patent ranking $R$:

$F = a \cdot R^{-b}$, where $a$ and $b$ are fitting parameters, and we simply set the values to a=1.0 and b=0.5. Similarly to the first step, the values of $F$ for each patent are normalized by the sum of $F$ values, and randomly generated $3n$ (the average number of forward citation per patent is three) values ranging between 0 and 1 are compared and assigned to the relative proportions of $F$ between 0 and 1.

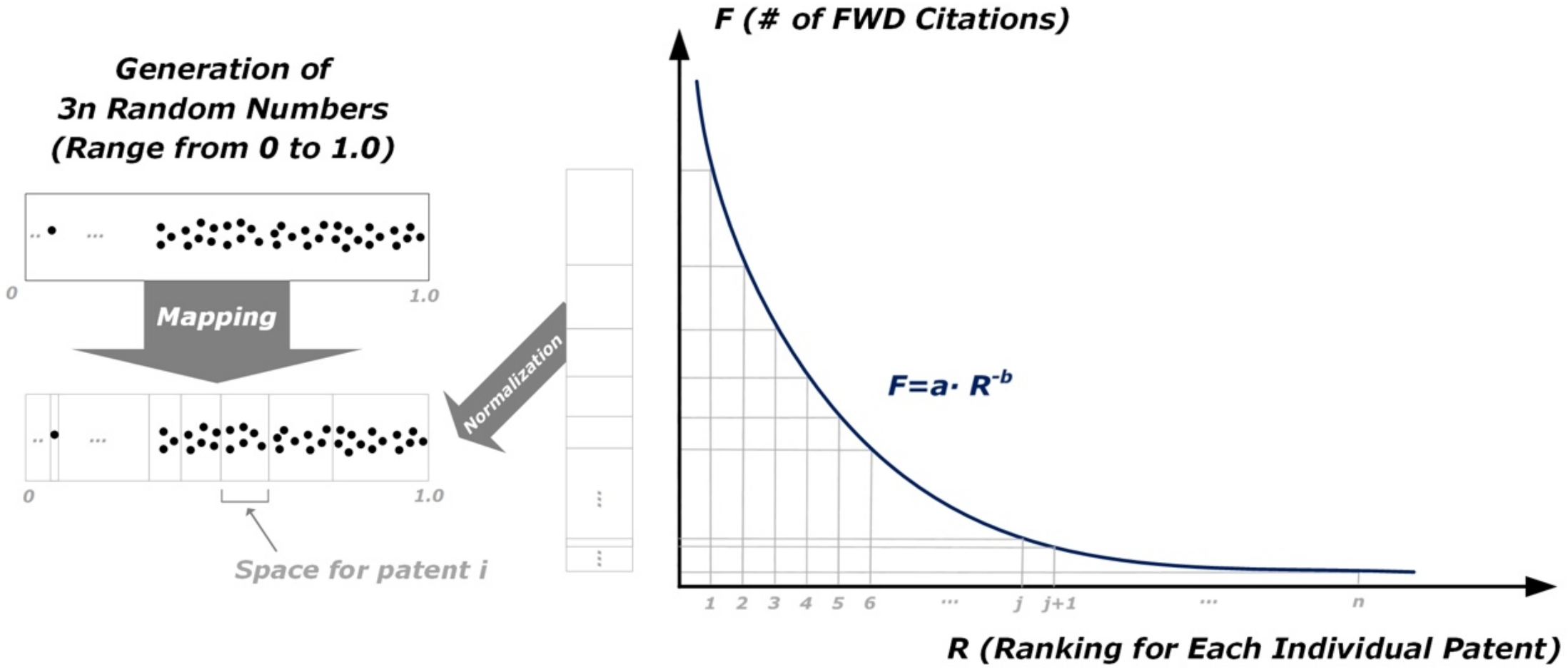

Fig 4. Number of assigned forward citation to each patent

Third, after assigning the number of forward citations of each patent, the years of the patents are determined (Fig 5). Since the number of patents for each year and the number of citations for each patent are decided, we assigned each patent to one of the available year-spaces by considering the number of forward citations and the number of late-occurring patents, i.e. potential citing patents. In particular, in order to fulfill the property of the citation lag, the whole time space after the year of a specific patent, i.e. *t+1~the last year* when the year of the patent is *t*, is divided into three periods, *5, 10 and whole timeframe*, and the different proportions of the number of forward citations are assigned to each period:

- 10% of citing patents of a specific patent are selected in period 1 (*t+1~t+5*),
- 10% of citing patents of a specific patent are selected in period 2 (*t+1~t+10*), except the patents already selected in period 1,
- The rest of citing patents of a specific patent are selected in period 3 (*t+1~the last year*), except the patents already selected in period 1 or 2.

Therefore, in order to assign a patent to a specific year *t*, the number of patents in each period has to be larger than the number of the assigned citations of the patent in a given period. When the available year-spaces of the patent are identified, the patent is randomly assigned to one of the available years.

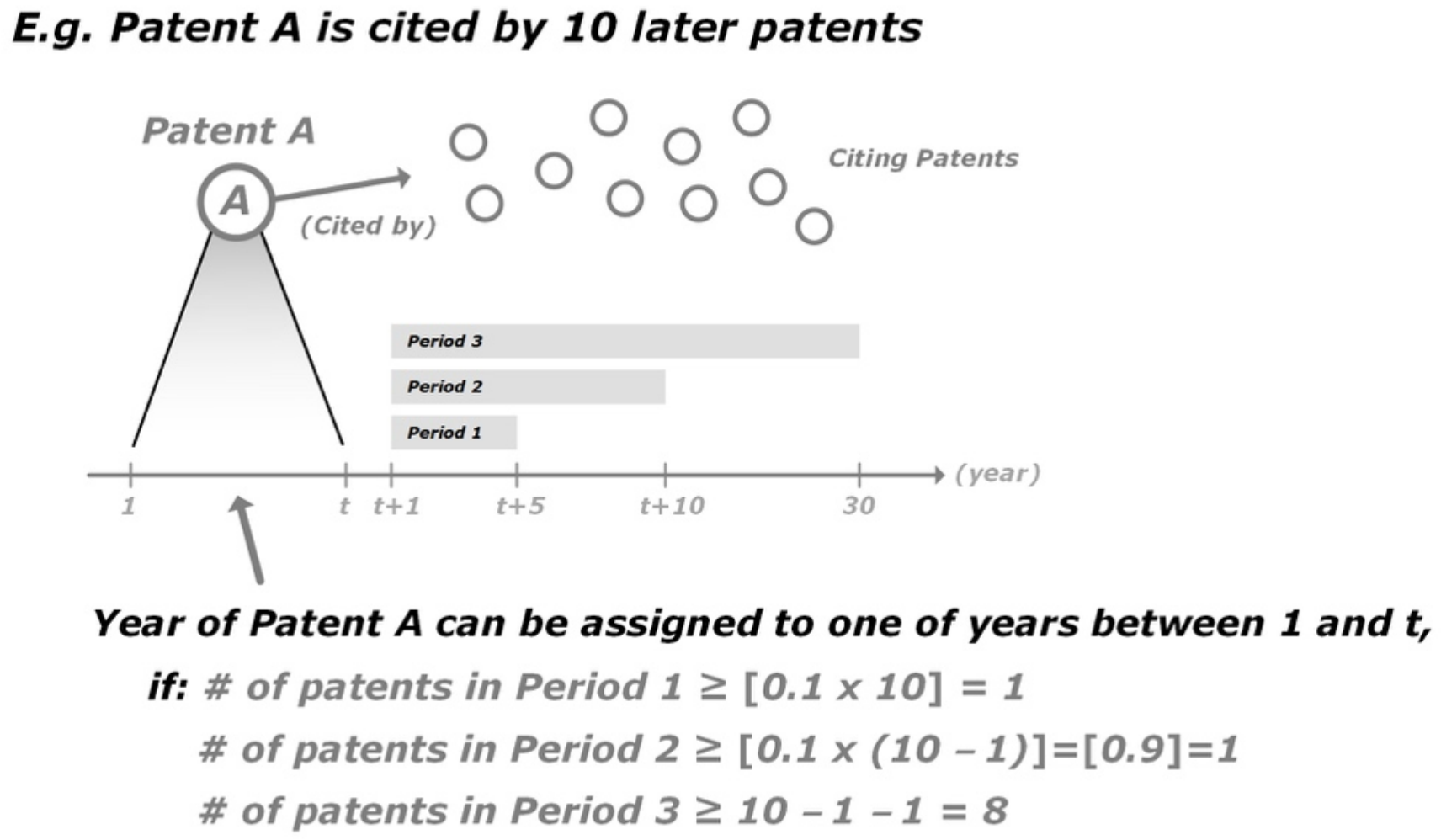

Fig 5. Allocation of patent's year

In the fourth step the specific forward citations are selected, i.e. the citing patents, of each patent (Fig 6). Based on the selection ratio for each period, the citing patents are randomly selected among the available patents without allowing duplication, i.e. if a patent in period 1 is selected as the citing patent, this patent cannot be selected again.

By this four step procedure, we generate all pairs of cited and citing patents and thus generate a patent citation network model. This network model is set by the total number of patents assumed to be in the domain being modeled but it is also stochastically determined and numerous variants of a particular model can be derived for a given number of patents in the domain. We run large number of samples to test the stochastic effects.

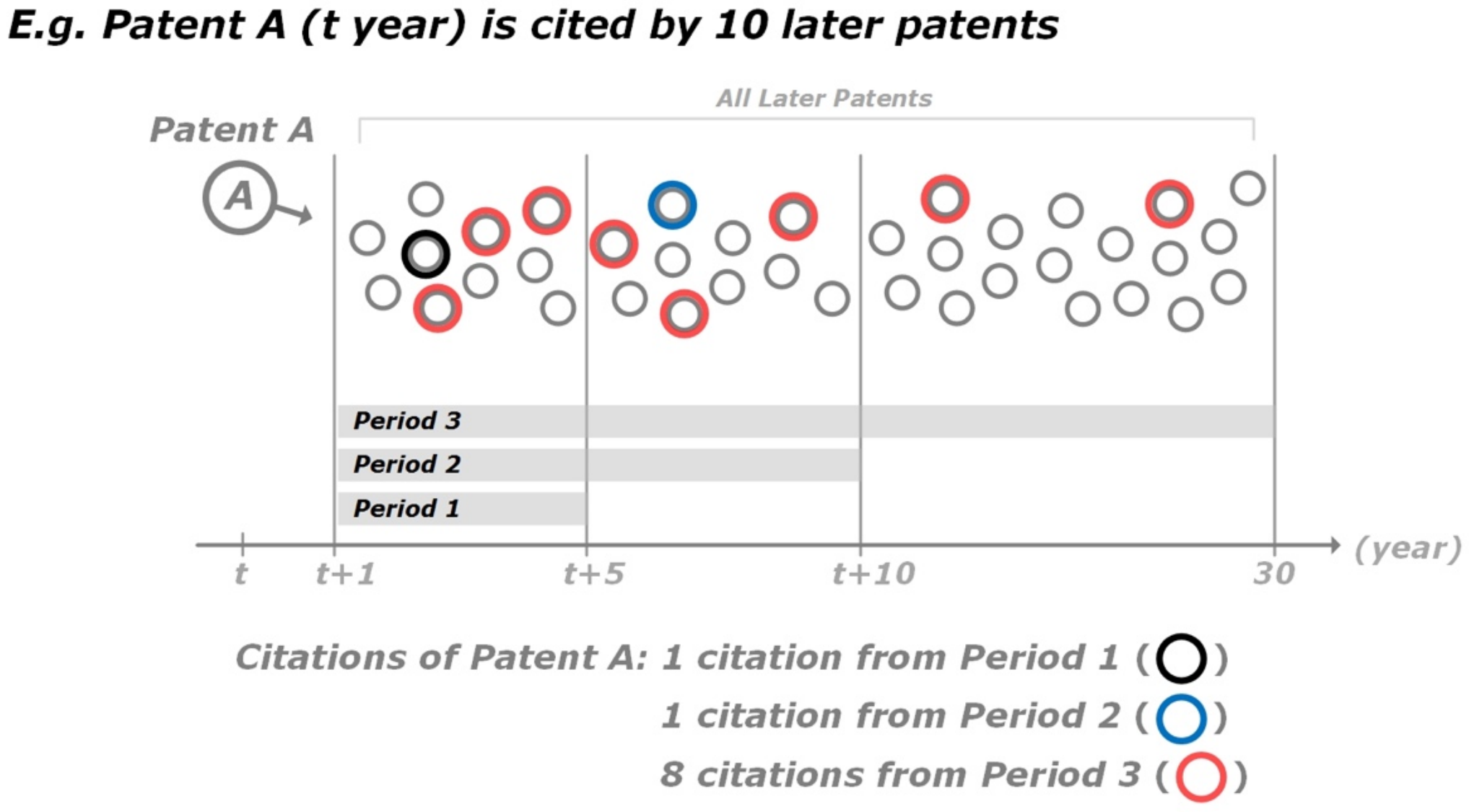

Fig 6. Selection of citing patents

### 3.2.2. Hypothetically designed discontinuity

The designed discontinuity plays a key role to test the proposed metrics. For our design goals, we specify that the designed discontinuity in a patent citation network is a patent which combines knowledge not previously combined and thereby introduces a new knowledge stream, and the inventive knowledge of the patent dominantly contributes to the later developmental process of the TD. In order to design a patent that meets these specifications, the different knowledge streams from two totally separate patent citation networks are first combined at the specific patent. In other words, before this patent, no patent in either one of these domains had cited a patent in the other domain. Cross-citation between other patents in the two domains between TDs is started from two years after the first combination and the proportion of combination increases over time, and then two originally different TDs are fully converged at the certain level of combination and considered as one new TD (the right hand side of Fig 7).

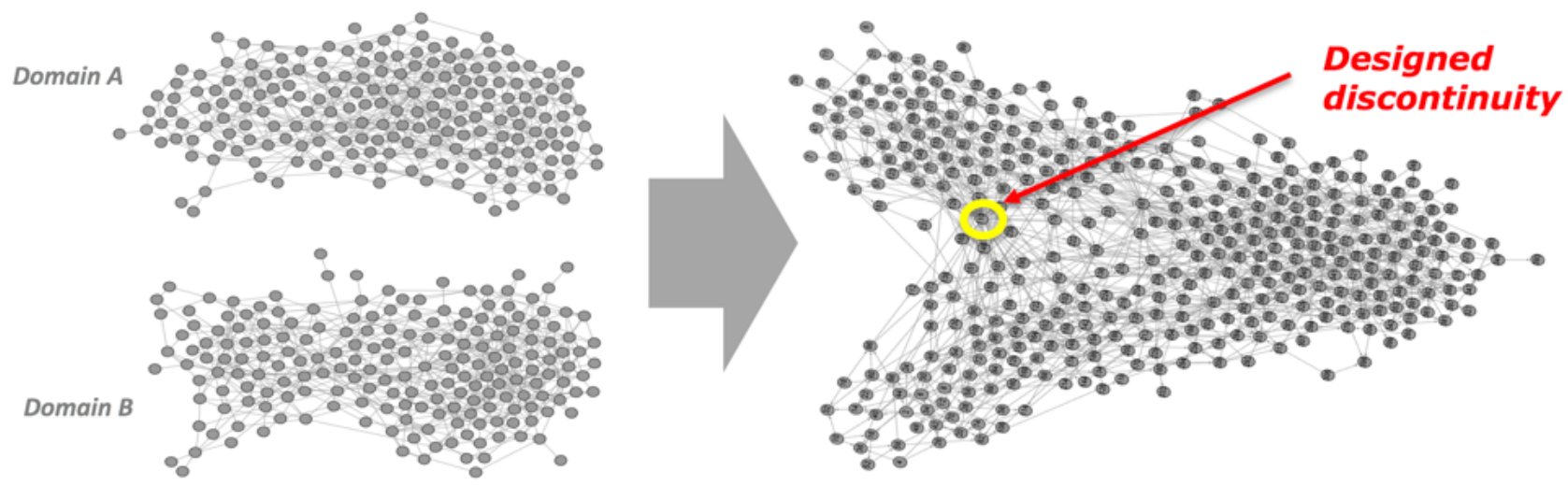

Fig 7. Creating a designed discontinuity from two previously unrelated domains

The procedure for generation of the designed discontinuity is as follows (Fig 8).

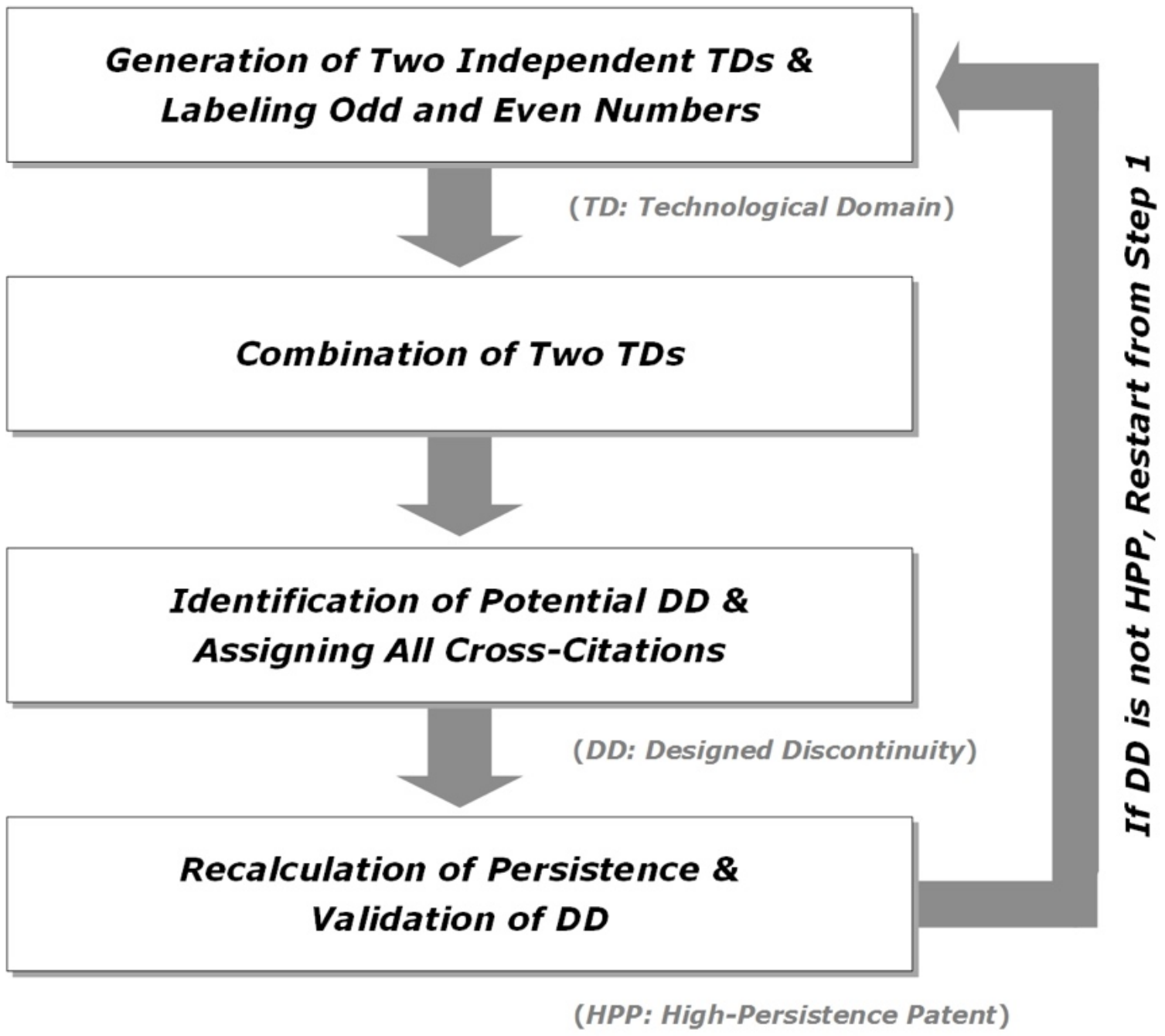

Fig 8. Procedure for generating designed discontinuity

First, two patent citation network models with 30-year period are generated and patents in one model are labeled by odd and in the other model by even numbers to make it easy to recognize their origin after the convergence.

Second, the two models are combined by cross-domain citations (Fig 9). Since patent citations denote knowledge flows, a constant increase of the occurrence frequency of cross-domain citations is used as a process of convergence to develop a new TD. The cross-domain citations are generated by replacing a backward citation of a patent in one domain with a citation by a patent in the other domain:

- when the patent is involved in *period l*, *k* % of backward citations of a patent (odd or even number) are replaced by patents of the other domain (even or odd numbers) in the same year with the original backward citation patent.

We do not change the number and original year of patent citations to maintain the generic characteristics of the patent model. In order to consider the increase of citation-replacement ratio, we set the different ratios for different periods:

- *k*=10%, when *period l* is 11~15 years,
- *k*=20%, when *period l* is 16~20 years,
- *k*=30%, when *period l* is 21~25 years,
- *k*=40%, when *period l* is 25~30 years.

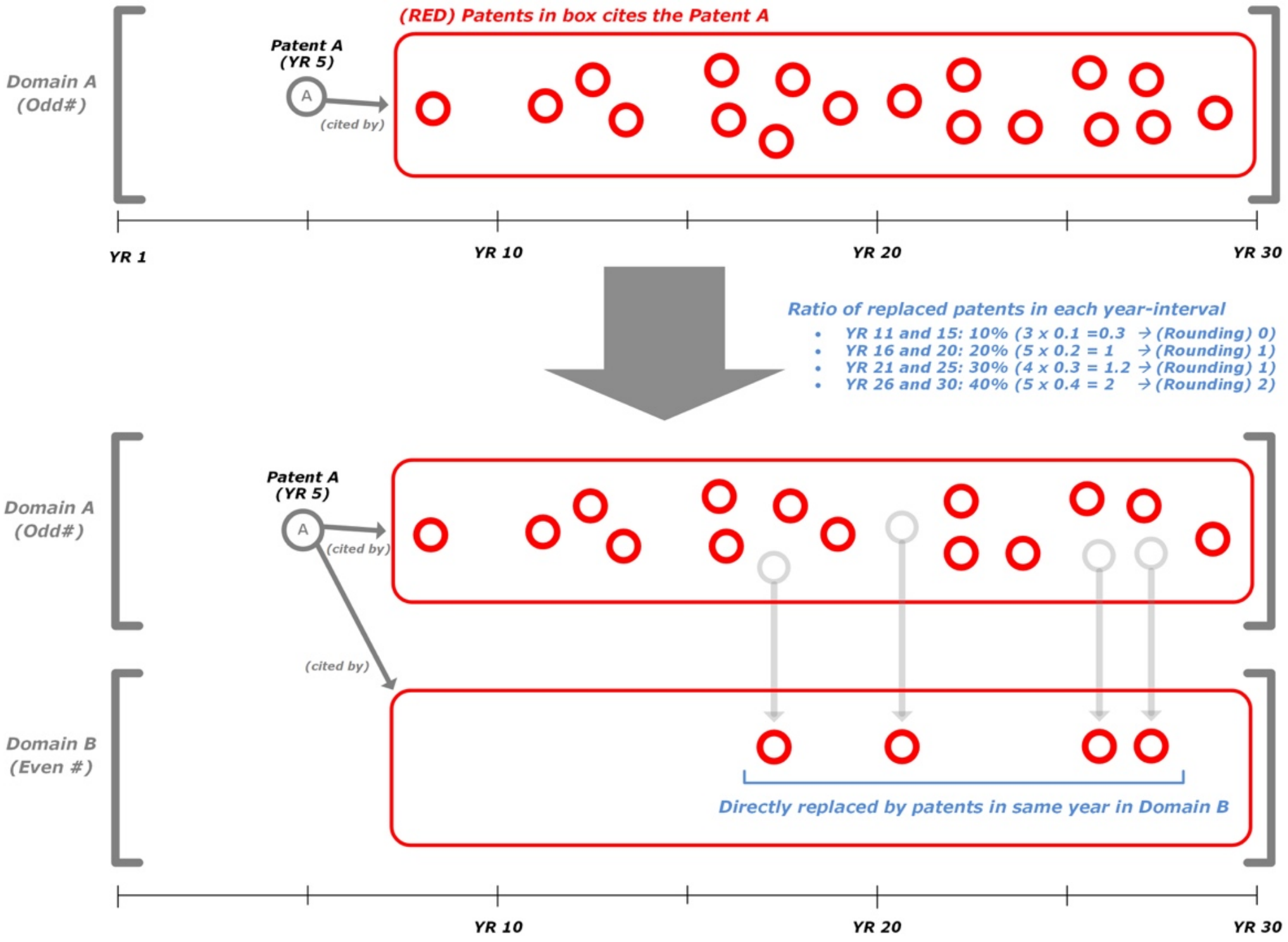

Fig 9. Process for generation of cross-domain citations

Third, in order to select the potential patent that can become the designed discontinuity, we identify a patent having the highest persistence value between year 11 and 13; thus, the designed discontinuity is created between year 11 and 13. All cross-domain citations occurring before the identified patent are replaced as backward citations of the selected patent to design it as the first patent that combines the two different knowledge streams.

In the last step, the persistence value of the focal patent is recalculated. If the patent has relatively high persistence value, i.e. HPP, after reallocating its backward citation structure, we identify the patent as the designed discontinuity; we considered a patent as the HPP when the normalized persistence[7] is over 0.7. However, if the normalized persistence of the patent is less than 0.7, the discontinuity generation process is restarted from the first step (Fig 8). Thus, our discontinuity achieves the design goals clearly combining previously separate knowledge streams and the status of high persistence patent that first combines the knowledge.

---

[7] Normalized persistence of a patent can be measured by dividing the highest persistence in the TD

## 4. Simulation

The test for the metrics is how well the metrics distinguish the designed discontinuity from other patents in the combined network model. In other words, if the metric reliably evaluates the designed discontinuity as the top-ranked patent, this metric is considered a potentially useful metric to identify knowledge discontinuities.

### 4.1. Reliability of simulation model

We first tested the reliability of the patent citation network model, and the result shows that the model closely follows the generic distributions of the patent system (Fig 10). In addition, even though we do not control the backward citation dynamics, the average backward citations in the model increases over time (Fig 10 (d) as do results from the actual patent system (Breitzman and Thomas, 2015)), our simulation is consistent with the aspect of the real world we want to reproduce.

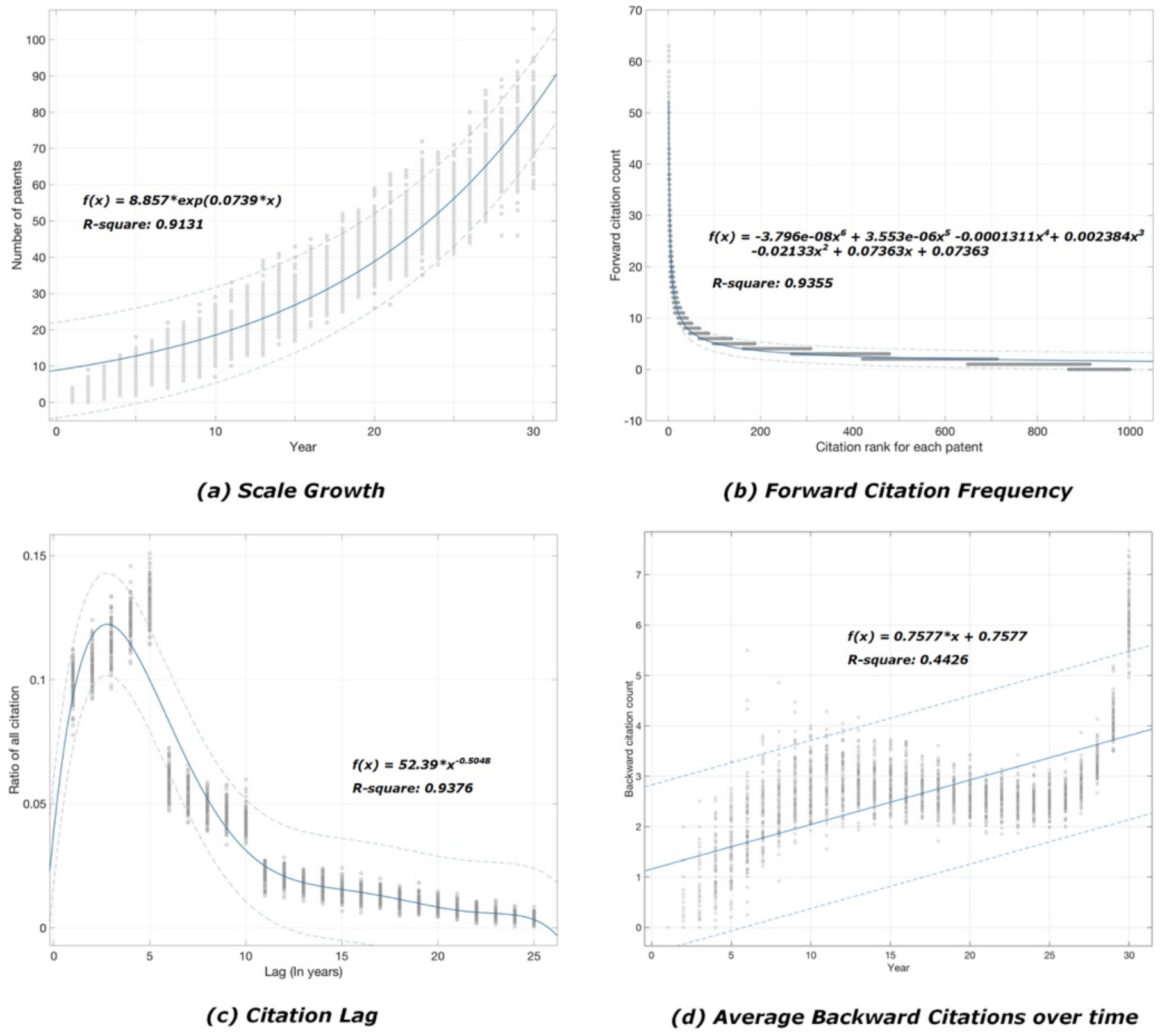

Fig 10. Distributional properties of simulation model. Note: 100 simulation tests was conducted and the size of each test was 1000; Blue line in each graph is a fitting curve and dotted lines denote 95% confidence bounds.

### 4.2. Simulation results

After assuring that the models in fact meet our intent, the value and ranking of the designed discontinuity for each simulation for each metric are calculated. To examine potential domain size effects, the combined network models were set with 600, 1,000, 2,000, 10,000, and 30,000 patents. To examine stochastic effects, each simulation was run 100 times. We set the performance criteria by six groups of top-ranked patents (Top 1, 3, 5, 10, 30 and 50) and evaluated the performance of the metrics by the probability of inclusion of the designed discontinuity in each group.

Each graph in Fig 11 shows how well the metrics can identify the designed discontinuity as one of top-ranked patents according to the size of a domain. For example, Metric 2 identifies the designed discontinuity as the top 1 patent out of 1000 patents with 28% probability (Fig 11 (b)).

The results show that Metric 5 performs best in this test. The main concept of Metric 5 is similar to the assumptions underlying the designed discontinuity, so this result was somewhat expected at the designing phase. Indeed, the results show that the performance of Metric 5 is clearly better than other metrics in any scale: Metric 5 can identify the designed discontinuity as the top ranked patent with 65~84% and as one of top 3 patents with 96~99% probability (Fig 11). In particular, Metric 5 seems to be the only metric that can reliably identify the designed discontinuity when the size of a domain is relatively small. In summary, Metric 5 (*persistence times # of converging paths*) should be carefully considered as the most important signal for identifying technological discontinuities. Metric 2 and 3 provide the second best performance from the simulation results; Metric 2 provides better performance than Metric 3 when the size of a domain is relatively small (Fig 11 (a) and (b)) but Metric 3 provides better performance when the size is relatively large (Fig 11 (d) and (e)). These second best metrics, however, are clearly inferior to Metric 5 with the performance gap between Metric 5 and the second best metric is considerable: the probability difference between Metric 5 and the second best metric ranges from 31% (Fig 11 (d)) to 49% (Fig 11 (c)).

Metric 4 is the worst performing of our metrics. Even though the concept of this metric, link disconnections of high-persistence patents on knowledge network, sounds plausible and Martinelli and Nomaler (2014) also used a similar concept to Metric 4 for identification of technological discontinuities, our simulation test shows Metric 4 does not reliably identify the designed discontinuity regardless of the size of a domain (Fig 11). The comparison of Metric 1 and Metric 2 shows that in the simulation, persistence provides a more significant signal than forward citations in terms of identification of knowledge discontinuities but that this superiority diminishes as size increases. Given that the designed discontinuities are the specific points where big changes in main knowledge occur and new dominant knowledge is introduced, it makes sense that a global metric like persistence, which follows successive inheritance of knowledge and is related to how much and how long specific knowledge makes impacts in the TD, is superior to a local metric such as forward citations.

Because of possible relationships between the design goals for the discontinuity and comparison of metric performance, a real world domain with a known discontinuity was also tested to examine the robustness of the simulation. We chose a domain where sufficient and deep qualitative work has been done previously and agreement exists about the patents associated with an important discontinuity. Our goal for this test case research is to probe the robustness of the simulation results. The simulation results to be compared with the test case include the inferiority of Metric 4, the better results with global (persistence) than local (forward citations) and the superiority of metric 5. Thus, we now turn to an empirical analysis of the magnetic information storage domain.

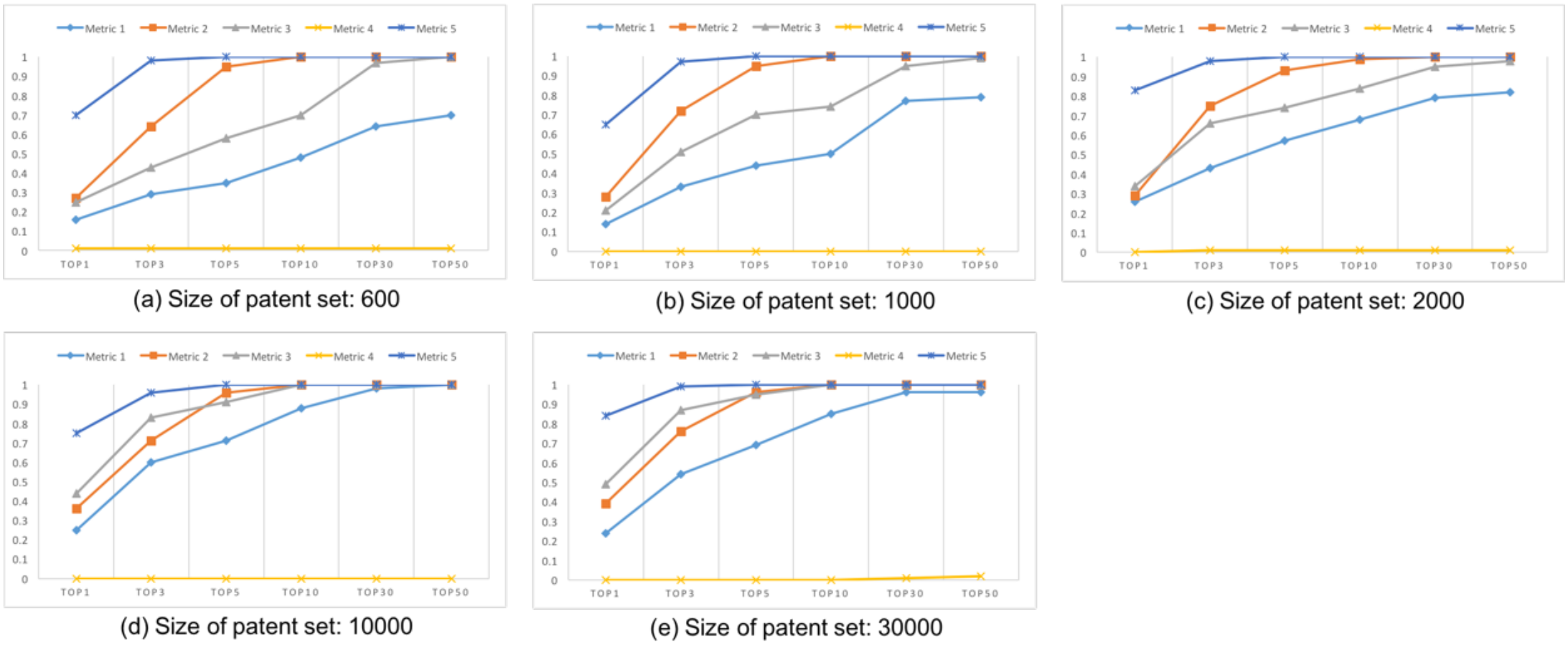

Fig 11. Simulation results. Note: *x*-axis designates the groups of top-ranked patents and *y*-axis is the identification probability in that group.

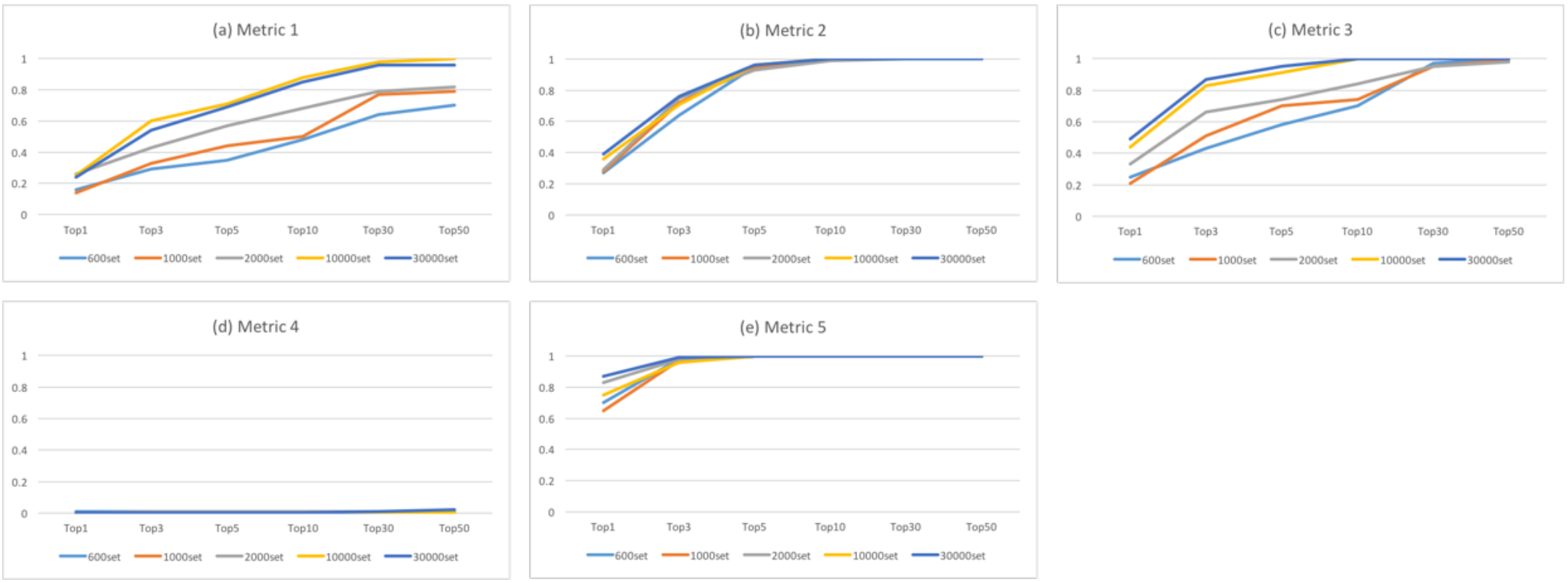

Fig 12. Sensitivity to scale. Note: y-axis is identification probability; x-axis is groups of top-ranked patents; line colors designate the size of a domain.

## 5. Empirical analysis: Magnetic information storage

### 5.1. Data

Although the simulation results allow a wide variety of sizes and time periods to be investigated, a further test using an actual set of patents was pursued to probe whether simulation and discontinuity design choices account for the findings. To examine the practical usability of the metrics and for comparison to the simulation results, we applied the metrics to a patent set for the magnetic information storage domain. Magnetic information storage is an appropriate case for this test in that the magnetic information storage is an important technological domain that make a great impact on information technology. Most importantly for our purposes, there is a clear knowledge discontinuity in its evolutionary pathway: the anisotropic magneto-resistance (AMR) head, which was the domain-standard head technology until the 1990s when a technological discontinuity – the GMR (giant magneto0resistance) head – also known as a spin-valve sensor – replaced it.

The GMR effect was discovered independently in 1988 by Albert Fert (University of Paris-Orsay) and Peter Grunberg (Julich Institute for Physics) and both received the Nobel Prize in Physics in 2007 for their independent discovery. Even though there were of course significant problems for application of a laboratory demonstrated GMR effect into an actual read head of a hard disk[8], GMR was quickly seen to have great potential to make improvements in the sensitivity of a read head, which can enable significant increases in an areal density. IBM was the first company that transformed the GMR effect from a laboratory demonstration into a working product[9]. IBM's GMR-based head sensor increased more than three times the areal density supported by previously dominant AMR heads,[10] and the rest of this domain adopted GMR heads in 1999.

We collected patents related to the magnetic information storage domain using the classification overlap method (COM)[11] from PatSnap (www.patsnap.com) and the summary of the patent set is shown in Table 1.

Table 1. Summary of patent set

| IPC and UPC overlap | Period (Application date) | The number of patents (Granted) | Relevancy |
|---|---|---|---|
| IPC (G11B) & UPC (360) | 1976.1.1 ~ 2013.7.1 | 33575 | 0.93 |

Extensive case based research has been done on the GMR effect and its translation to important technological results (Bajorek, 2014; Cros et al., 2009; Dieny, 2007; Sharpe, 2009; Thompson, 2008). Examination of this literature indicates that three patents for GMR heads (US 4949039, US 5206590 and US 5159513) are representative of this technological discontinuity. Summaries of the patents are shown in Table 2.

Table 2. Summary of technological discontinuity patents

| Patent number | Title | Application date | Assignee | Inventors | Reference |
|---|---|---|---|---|---|
| US 4949039 | Magnetic field sensor with ferromagnetic thin layers having magnetically antiparallel polarized components | Jun 14, 1989 | Kernforschungsanlage Julich Gmbh | Peter Grunberg | Bajorek (2014); Cros et al. (2009) |

[8] The GMR effect was first discovered using specimens produced by the very expensive Molecular Beam Epitaxy (MBE) method and extremely high magnetic fields under very low temperatures (-449 degrees F) in 1988. However, for commercialization, the GMR effect had to be generated at room temperatures with small magnets by an affordable manufacturing method.

[9] IBM announced the first GMR head-based 3.5 inch 16.8 gigabytes hard disk 'IBM Deskstar 16GP DTTA-351680' in 1997.

[10] In the middle and late 1990s, AMR heads supported the areal densities as high as 3.3 gigabits per square inch, but GMR heads supported the areal densities greater than 10 gigabits per square inch. Since the exponential rate of improvement in this domain is large (~35% per year), such an increase is not statistically surprising. Nonetheless, measurement over times show a clear break to an increased exponential slope of the areal density for magnetic storage in the mid-90's (Magee et al., 2014).

[11] The classification overlap method (COM) is a patent search technique, developed by Benson and Magee (2013, 2015b) that can isolate sets of patents for TDs. The main idea of the COM is that every US patent is classified by experts within two different patent classification system, US Patent Classification (UPC) and International Patent Classification (IPC), and patents listed in *both* specific UPC and IPC are shown to be a highly relevant and complete set. The relevancy between the patents and the TD is on average 86% (97% for the best and 67% for the worst case) by testing 28 TDs.

| | | | | | |
|---|---|---|---|---|---|
| US 5206590 | Magnetoresistive sensor based on the spin valve effect | Dec 11, 1990 | IBM | Bernard Dieny, Bruce A. Gurney,<br>Steven E. Lambert, Daniele Mauri,<br>Stuart S. P. Parkin, Virgil S. Speriosu,<br>Dennis R. Wilhoit | Dieny (2007); Bajorek (2014); Cros et al. (2009); US 5159513 |
| US 5159513 | Magnetoresistive sensor based on the spin valve effect | Feb 8, 1991 | IBM | Bernard Dieny, Bruce A. Gurney,<br>Serhat Metin, Stuart S. P. Parkin,<br>Virgil S. Speriosu | Sharpe (2009); |

First, US patent 4949039 invented by Peter Grunberg, a recipient of the Nobel Physics Prize for GMR in 2007, is about a layered magnetic structure that generates an enhanced magneto-resistance effect, i.e. GMR, caused by the anti-parallel magnetic alignment of the magnetic layers and is almost the first patent on generating GMR effect and seen as the basic key patent for magneto-electronics (Cros et al., 2009). When he discovered the GMR effect, Grunberg realized the high potential of application of the GMR effect to AMR sensors, because the structure of the GMR sensor was easily compatible with the AMR sensors (Binasch et al., 1989; Thompson, 2008) and applied for a patent. This patent has been acknowledged by all major firms in the magnetic information storage domain and earned more than 10 million dollars in licensing fees (Dedrick and Kraemer, 2015).

Second, US 5206590 and US 5159513 filed by Stuart Parkin's group at IBM's Almaden Research Center are recognized to be the first two patents on the spin-valve sensors, which critically affect almost all further GMR head technologies by IBM and others (Bajorek, 2014; Sharpe, 2009). Specifically, US 5206590 is about the basic structure and concept of the spin-valve sensor, filed in 1990 (Cros et al., 2009; Dieny, 2007), and US 5159513 is about the preferred materials for the ferromagnetic layers, cobalt or cobalt alloy, based on the basic spin-valve structure.

## 5.2. Empirical results and discussion

The performance of Metric 5 is tested by how well the metric can identify and distinguish the aforementioned three patents from others. The empirical results in Fig 13 show that Metric 5 successfully identifies the three patents as top-ranked discontinuities (ranking 2, 5 and 11) out of more than 30,000 patents (Table 3). Furthermore, our qualitative investigation of the patents that are highly ranked by Metric 5 (Table 4) reveals that Metric 5 identifies not only the representative GMR related discontinuity in 1989~1991, but also other important discontinuities in different timeframes. US patent 4103315 (7th) is IBM's and the domain's first MR head technology and is the technological foundation of further MR and GMR related inventions and has been generally recognized as a technological discontinuity occurring in the late 1970s; US patent 4663685 (3rd) and 4535375 (4th) are about a MR read sensor technology that can eliminate Barkhausen noise which was a critical bottleneck in a MR head sensor in the early 1980s. Therefore, Metric 5 provides reliable performance not only from the simulation test but also from the empirical test.

The empirical results in Table 3 also show that Metric 4 does not identify the GMR knowledge discontinuity since none of the three patents are in the top 15 identified by this metric and two of the patents are not identified at all. The combination of the simulation results and this empirical case study support the conclusion that this metric is inappropriate for use in detecting knowledge discontinuities. Thus, the idea that weakly linked persistent patents (our framing of the well regarded cyclic pattern – Abernathy and Utterback, 1978; Anderson and Tushman, 1990; Kaplan, 1999; Munir, 2003; Sahal, 1981;

Tushman, 1997; Tushman and Nelson, 1990; Utterback, 1994) are important aspects of technological discontinuities gets no support from our simulation or empirical results. The second weakest metric in this case is Metric 1 which was also the next weakest in the simulation results. While Metric 1 (*forward in-domain citations*) did identify one of the discontinuity patents as its top ranked patent, it was weak in the other two patents which indicates that a global measure such as persistence is in fact a better way to identify knowledge discontinuities as originally argued theoretically by Martinelli and Nomaler (2014). The overall agreement of the simulation results and the magnetic information storage results strengthens the conclusion that technological discontinuities are not always top-cited patents.

Even though the second and third best metrics (Metric 2 and 3) also successfully identifies the three discontinuity patents (Table 3) and appear to give adequate results overall, use of Metric 5 is recommended. Metric 5 has no drawbacks and is clearly the best metric in the simulation test. Overall, the empirical case study aligns with the simulation to the full extent possible since the qualitative differences seen in the case for Metric 1 and 4 reaffirm the simulation results and no qualitative analysis can establish any other metric as superior to metric 5. Qualitative examination of a TD can identify some important technological discontinuities, such as GMR head and MR head, but it is not possible to objectively specify the ranking among them without a metric and in these results, Metric 5 is clearly best overall.

Table 3. Summary of results

| Patent number | Metric 1 (Normalized) (Ranking) | Metric 2 (Normalized) (Ranking) | Metric 3 (Normalized) (Ranking) | Metric 4 (Normalized) (Ranking) | Metric 5 (Normalized) (Ranking) |
|---|---|---|---|---|---|
| US 4949039 | 59<br>(0.301)<br>(76) | 369.10<br>(0.896)<br>(2) | 21776.95<br>(0.676)<br>(5) | 1.05<br>(0.003)<br>(17) | 123.03<br>(0.747)<br>(2) |
| US 5206590 | 196<br>(1.0)<br>(1) | 164.25<br>(0.399)<br>(10) | 32192.88<br>(1.0)<br>(1) | N/A<br>(N/A)<br>(N/A) | 41.06<br>(0.249)<br>(11) |
| US 5159513 | 102<br>(0.52)<br>(19) | 169.56<br>(0.412)<br>(9) | 17294.93<br>(0.537)<br>(6) | N/A<br>(N/A)<br>(N/A) | 67.82<br>(0.412)<br>(5) |

* Normalized value is a proportion of the maximum value of each metric.

Table 4. Top 15 patents by Metric 5

| Ranking by Metric 5 | Patent number | Application date | Title |
|---|---|---|---|
| 1 | US4755897 | 1987 | Magnetoresistive sensor with improved antiferromagnetic film |
| 2 | US4949039* | 1989 | Magnetic field sensor with ferromagnetic thin layers having magnetically antiparallel polarized components |
| 3 | US4663685 | 1985 | Magnetoresistive read transducer having patterned longitudinal bias |
| 4 | US4535375 | 1983 | Magnetoresistive head |
| 5 | US5159513* | 1991 | Magnetoresistive sensor based on the spin valve effect |
| 6 | US5287238 | 1992 | Dual spin valve magnetoresistive sensor |
| 7 | US4103315 | 1977 | Antiferromagnetic-ferromagnetic exchange bias films |
| 8 | US4280158 | 1979 | Magnetoresistive reading head |
| 9 | US4356523 | 1980 | Narrow track magnetoresistive transducer assembly |
| 10 | US4879619 | 1988 | Magnetoresistive read transducer |
| 11 | US5206590* | 1990 | Magnetoresistive sensor based on the spin valve effect |
| 12 | US4130845 | 1977 | Disc cabinet recirculating air flow system |
| 13 | US5005096 | 1988 | Magnetoresistive read transducer having hard magnetic shunt bias |
| 14 | US4419701 | 1981 | Data transducer position control system for rotating disk data storage equipment |
| 15 | US6317297 | 1999 | Current pinned dual spin valve with synthetic pinned layers |

** : The three patents associated with the GMR discontinuity.*

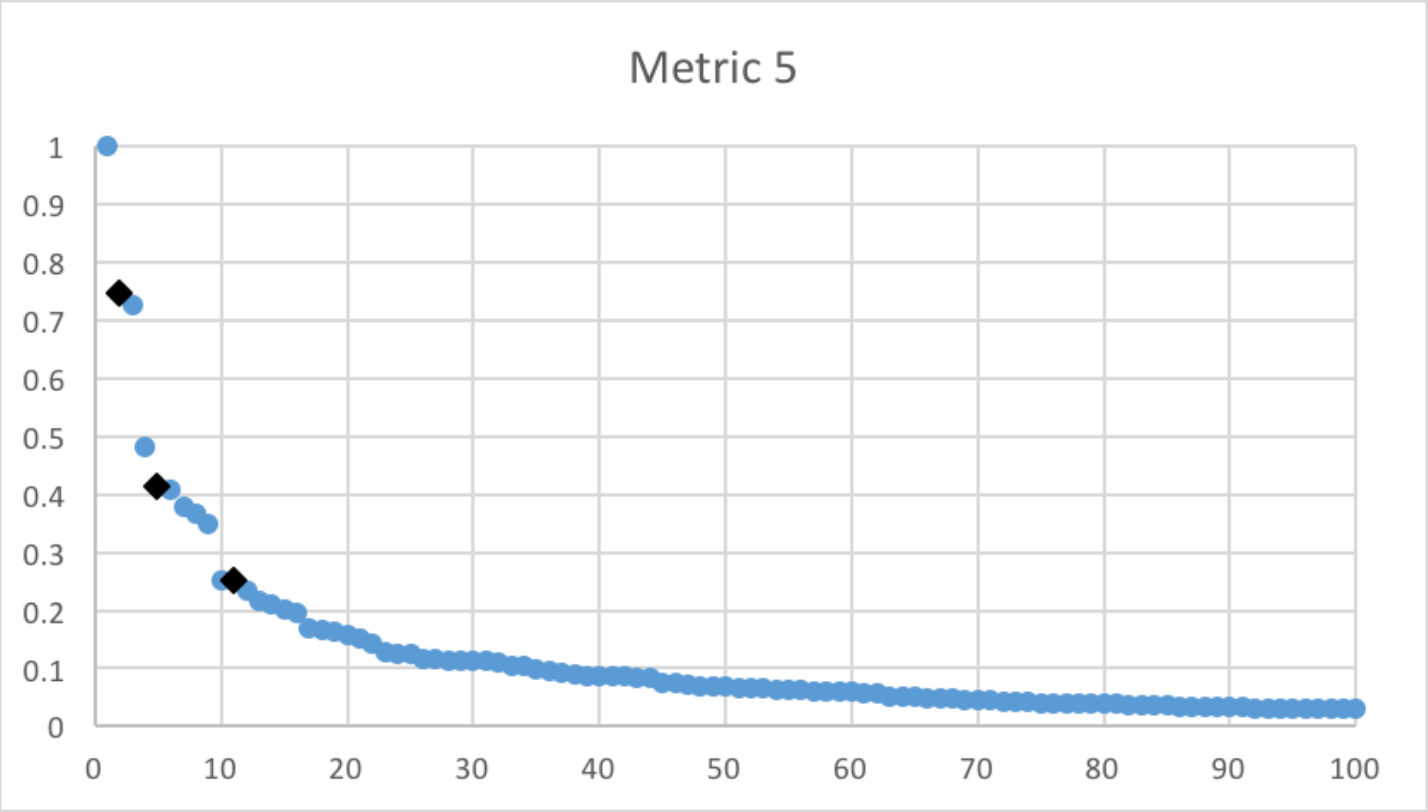

Fig 13. Normalized value of top 100 patents by Metric 5. Note: black diamonds are the three knowledge discontinuities; x-axis is top 100 patents by each metric and y-axis is the normalized value

## 6. Conclusions

This paper proposes metrics to quantitatively identify technological discontinuities in a knowledge network of technological domains. We developed five metrics and tested them by developing a simulation model-based knowledge network containing a hypothetically designed discontinuity. The designed discontinuity was developed as a specific patent that converges two totally different knowledge streams and the major knowledge of the patent is dominantly persistent in the knowledge network. The performances of the metrics are evaluated by how well the metrics can identify the designed discontinuity as one of top-ranked patents. The simulation results show that the reliabilities of two simulation models, a patent citation network model and combined network model, are effective in following the basic distributional characteristics of the real world patent system. The three metrics, persistence, persistence times forward citations, and persistence times # of converging paths, provide good performance in identifying the designed discontinuity with persistence times # of converging paths consistently showing the best performance. The simulation also shows that forward citations and particularly minimum persistence do not identify the discontinuity well. We also conducted an empirical test using a patent set for the magnetic information storage technological domain. The three technological discontinuity patents were selected based upon prior, extensive qualitative studies and the performance of the metrics was assessed by how well the metrics identify the three selected patents as top-ranked patents out of the 30,000+ patents in this technological domain. The empirical results agree with the simulation results and support that persistence times # of converging paths is the appropriate metric to identify technological discontinuities in a TD. The simulation results indicate that domain sizes between 600 and 30,000 patents do not have a strong effect on the results so applying the metric to a variety of domains is appropriate.

However, there exist several issues to resolve in further research. The first issue arises since the major focus of this research is to develop the metrics and testing the performance of them by a simulation model to examine conceptual robustness of the metrics. Even though this research conducted an empirical test for one specific domain, it is not sufficient to fully define application procedures and other issues. Therefore, an important next step for further research is an actual application of the metrics to many technological domains. We suggest that this work focus on use of Metric 5 to identify top candidates for discontinuities in each domain and then qualitatively examine the identified patents and do some comparisons to the other metrics. Such work could greatly increase our knowledge of discontinuities in a number of domains as well as define better procedures for using the metrics in their identification. The second issue is that we did not attempt to develop new or better metrics since three of our metrics showed good performance but further effort along this path might be worthwhile. The third important issue recognizes that this research and all path finding research uses patent forward citations; therefore it will be difficult to identify the very recent technological discontinuities. Thus new approaches for identifying future technological discontinuities will be an important topic for research.

**Acknowledgement**

The authors acknowledge the funding support for this work received from the SUTD-MIT International Design Center and the Hanyang University (HY-2016).